\documentclass[aps,prl,twocolumn,draft
showpacs,floatfix,amsfonts,amssymb]{revtex4}
\usepackage{graphicx,graphics,color,epsfig}
\usepackage{bm}
\usepackage{amsmath}
\usepackage{amssymb}
\usepackage{amsfonts}

\newcommand{\br}{{\bf r}}

\newcommand{\beqa}{\begin{eqnarray}}
\newcommand{\eeqa}{\end{eqnarray}}

\begin{document}

\title{Effect of impurities on supersolid condensate: a
Ginzburg-Landau approach
}
\author{Alexander V. Balatsky$^1$ and Elihu Abrahams$^2$}
\affiliation {$^1$Theoretical Division, Los Alamos National
Laboratory, Los Alamos, New Mexico 87545 \\
$^2$Center for Materials Theory, Serin Physics Laboratory, Rutgers
University, 136 Frelinghuysen Road, Piscataway, New Jersey 08854}
\date{\today}
\begin{abstract} Basing our arguments on a wave function that
contains both positional and superfluid order, we propose a
Ginzburg-Landau functional for a supersolid with the two order
parameters necessary to describe such a  phase: density $n_B(\br)$
and supersolid order parameter $\psi_V$. We argue that adding
lighter $^3$He atoms to a $^4$He supersolid produces attractive
regions for vacancies, leading to patches of higher $T_c$.  On the
other hand, the supersolid stiffness decreases in this granular
state with increased $^3$He disorder. Both effects are linear in
$^3$He concentration.
\end{abstract}
\maketitle
\section{Introduction}Recent experiments by Chan and Kim
\cite{Chan04, Chan:05} have generated renewed interest in the
possibility of the existence of a supersolid phase in $^4$He.
There is pioneering  theoretical work by, among others,  Andreev
and Lifshitz \cite{AL}, Reatto \cite{Reatto67}, Chester
\cite{Chester70}, Leggett \cite{Leggett70} and Anderson
\cite{Anderson84}. Recently, Anderson \cite{Anderson05} and
Anderson, Brinkman and Huse \cite{ABH} have discussed the problem.
Recent developments have been reported at a KITP workshop and are
available online \cite{KITP}

The main purpose of purpose of this note is to develop an approach
that allows a discusssion of the effect of $^3$He
impurities on a $^4$He supersolid. In what follows, we show how
the approach leads to a Ginzburg-Landau functional for the
supersolid. We propose to  follow the collective degrees of
freedom  that are candidates for the supersolid properties, namely
the vacancies  that are intrinsic to an incommensurate
\cite{ABH} solid. Thus, we follow Andreev and Lifshits \cite{AL},
and more recently Anderson
\cite{Anderson05}, who argued that the  supersolid requires some
defects, which are inherently present in the phase at $T=0$.
Vacancies and interstitials are obvious candidate defects in the
$^4$He solid. For simplicity, we focus here on vacancies
\cite{activ}. In this picture, the lattice sites are equivalent,
there is no disorder and yet the He solid is not commensurate.

Thus, we can consider the ground state of $^4$He at $T=0$ as a
state in which there is a non-zero  concentration of vacancies.
Although a localized vacancy may require a non-zero activation
energy, the fact that they are mobile (the more so due to the
large zero-point motion of the He atoms) can bring the bottom of
their band to zero energy. For any non-zero concentration $n_V$ of
vacancies, they will form a Bose condensate at $T=0$. However, to
properly describe the quantum nature of the condensate we have to
address not the densities but the quantum amplitudes of the
relevant degreees of freedom. Thus the Ginzburg-Landau (GL)
functional for the supersolid deals with the vacancy Bose field
$\psi_{V}= \sqrt{n_{V}}
\exp(i\alpha)$ as an order parameter.


\section{Wave function for a supersolid} In this section, we
discuss wave functions that display solid order with,  at the same
time, a Bose condensed fraction of vacancies.

We start with a lattice state of hard-core bosons; the boson
occupancy cannot exceed one per lattice site. The operator $b_i^+$
creates a $^4$He atom at site
$i$ of the lattice. The commensurate crystal wave function is then
\beqa |\Psi_0\rangle =   \prod_i b_i^+|0\rangle =|\uparrow,
\uparrow, ...\uparrow \rangle,
\eeqa  corresponding to a ``ferromagnetic" state, with no
vacancies. For some purposes, it is convenient to use a spin-1/2
representation \cite{zilsel}:
\beqa S^+_i = b^+_i, \  S^-_i = b_i, \ S^z_i = b^+_ib_i - 1/2.
\eeqa To introduce vacancies, a simple procedure would be to apply
$\sum_i S_i^-$ $N_V$ times to create $N_V$ vacancies. However, it
has been understood since BCS and was pointed out in the present
context by Anderson
\cite{Anderson05} that a phase coherent superfluid state must be a
linear combination of states of different particle number. To
achieve this, we apply a tilt operator to the ``ferromagnetic"
commensurate state. The rotation operator in the spin
representation is (now using Pauli operators)
\beqa \prod_i \exp(i {\bm \sigma}_i \cdot  {\hat{\bf a}}_i
\phi_i/2) =
\prod_i[\cos {1\over 2}\phi_i + i ({\bm \sigma}_i\cdot{\hat{\bf
a}_i})
\sin{1\over 2}\phi_i].
\label{EQ:twist1}
\eeqa Here, $\phi_i$ is the tilt angle at the $i$-th site and
${\hat{\bf a}}_i$ is the rotation axis, which lies in the $x,y$
plane. The latter's angle with respect to the $y$ axis will be
denoted by
$\alpha_i$. Since ${\hat{\bf a}}_i \perp {\hat{\bf z}}$,
\beqa {\bm\sigma}_i\cdot {\hat{\bf a}}_i  = (S_i^+ a_i^- + S_i^-
a_i^+).
\eeqa With $a_i^{\pm} = a_x \pm ia_y = \pm i \exp (\pm
i\alpha_i)$, we get
\beqa {\bm\sigma}_i\cdot {\hat{\bf a}}_i  = i e^{i\alpha_i} b_i - i
e^{-i\alpha_i} b_i^+.
\eeqa

We operate with Eq.\ (3) on $|\Psi_0\rangle = \prod_i
b_i^+|0\rangle$ and use Eq.\ (5) to find the tilted state \beqa
|\Psi\rangle &=&
\prod_i [\cos{1\over 2}\phi_i +i(ie^{i\alpha_i})\sin{1\over
2}\phi_i
\ b_i ] b_i^+ |0\rangle  \nonumber\\ & = & \prod_i[\cos{1\over
2}\phi_i \ b_i^+ - e^{i\alpha_i}\sin{1\over 2}\phi_i]|0\rangle,
\label{EQ:twist2}
 \eeqa
  where we have used the fact that because of the hard-core
constraint, $S_i^+ = b_i^+$ gives zero acting on
$|\Psi_0\rangle$. The state $|\Psi\rangle$ is a linear combination
of states with specified phases and different numbers of
vacancies; it is normalized and the expectation value of vacancy
occupation is:
\beqa n_V^i  = 1- \langle b_i^+b_i\rangle = \sin^2{1\over 2}\phi_i
\eeqa The vacancies are mobile due to rearrangements of the atoms,
which is facilitated by their large zero-point motion. We model
the kinetic energy of vacancies with a simple nearest-neighbor
hopping hamiltonian \cite{multi}. Its expectation value in
$|\Psi\rangle$ is \cite{KE}: \beqa
 KE &=& -t\sum_{\langle ij\rangle}\langle b_i^+b_j + h.c.\rangle
\nonumber\\ & = &(-t/2)
\sum_{\langle ij\rangle}\sin \phi_i \sin
\phi_j\cos(\alpha_i-\alpha_j).
\eeqa

With all ${\phi_i,\alpha_i}$ the same, $|\Psi\rangle$  is a state
with both solid order and  a phase coherent vacancy contribution.
The vacancy order parameter is
\beqa
\psi_V= \langle b_i\rangle ={1\over 2} \sin \phi e^{-i\alpha}.
\eeqa

In the spin language, when all ${\phi_i,\alpha_i}$ are the same,
$|\Psi\rangle$ is a state with maximal total spin $S=M/2$, where
$M$ is the number of lattice sites. The expectation value of
$S^z = \sum S_i^z$ is $\langle S^z\rangle = (M/2)\cos \phi$.

\section{Ginzburg-Landau Functional}

We start with with the postulate that  the GL functional of a
supersolid phase in $^4$He has to contain {\em two} order
parameters: density $n_B(\br)$ and superfluid amplitude
$\psi_V(\br)$. Although there are only
$^4$He atoms, these atoms particpate in two distinct phenomena:
solid and superfluid. In the region of the experimental phase
diagram \cite{Chan04}, where one enters a possible supersolid
phase from the solid at $ T \sim 100 mK$ at fixed pressure, solid
order is presumably already well established and hence is well
outside of the GL regime. Thus, we concern ourselves with a
possible second-order transition across a normal solid to
supersolid transition line. The amplitude
$|\psi_V(\br)|$  is small at the transition and is in the GL
regime. The wave function $|\Psi\rangle$ in Eq.\ (\ref{EQ:twist2})
offers a unified description of  both solid order with
periodically modulated density $n_B(\br)$ and vacancies with the
order parameter
$\psi_V$ plus the constraint $n_B(\br) + n_V(\br) = 1$. In the
discussion below we focus on the $\psi_V$ field.

Toward developing the GL approach, we assume that the vacancy
density is small (as is consistent with the experimental
situation) so that the tilt angle $\phi$ is small. To estimate the
superfluid stiffness, we examine the $KE$ as the phase
$\alpha$ acquires a slow variation from site to site. From Eqs.\
(8,10)and small $\phi$, we get
\beqa
\Delta\, KE = (t / a_0)\int d\br |\psi_V|^2 (\nabla\alpha)^2,
\eeqa where
$a_0$ is the lattice spacing in a simple cubic lattice. We see
that the superfluid stiffness  is determined by the hopping
amplitude $t$ and the density of vacancies $|\psi_V|^2$.

Next we examine the effect of a potential that couples to the
$^4$He density,
$H_1=U_i b_i^+ b_i$. From Eq.\ (7),
\beqa
\langle H_1\rangle & = & \sum_i U(i) \cos^2{1\over 2}\phi_i
\nonumber\\ &
\approx &  - (1/\Omega)\int d{\bf r} \,u({\bf r}) \sin^2{1\over
2}\phi({\bf r})
\nonumber\\ & \approx & - (1/\Omega)\int d\br \,u(\br)
|\psi_V(\br)|^2,
\eeqa where $\Omega$ is the unit cell volume,
$u(\br) = \Omega \sum_i U_i \delta(\br -\br_i)$ is the potential
energy  and we ignored the constant $\int d\br\, u(\br)$.  In
the last line, we took advantage of the approximation that $\phi$
is small so that, from Eq.\ (10),
$|\psi_V|^2 \approx n_V = \sin^2{1\over 2}\phi $. In our proposed
state
$|\Psi\rangle$, a potential that is {\em repulsive} for $^4$He
atoms would be {\em attractive} for vacancies, tending to increase
their local density.

Our discussion leads to the total GL free energy density
\beqa F(\br) &=& [a({T\over T_c^0}-1) -
u(\br)/\Omega]|\psi_V(\br)|^2  + (t/a_0)|\psi_V(\br)|^2|\nabla
\alpha|^2 \nonumber\\
 &+ &b |\psi_V(\br)|^4.
\eeqa The quartic term in $F$ represents the restriction against
putting two vacancies in the same place, as well as dynamical
interaction terms.

\section{Effect of disorder} Chan et al \cite{Chan:05} find the
remarkable result that by adding $^3$He atoms in the $^4$He solid,
the measured $T_c$ grows and the superfluid stiffness $\rho_s$
drops as a function of
$^3$He concentration. The data is reproduced here in Fig.\ 1.
\begin{figure}[t]
\begin{center}
\vskip -.4in
\includegraphics[width=8.5cm]{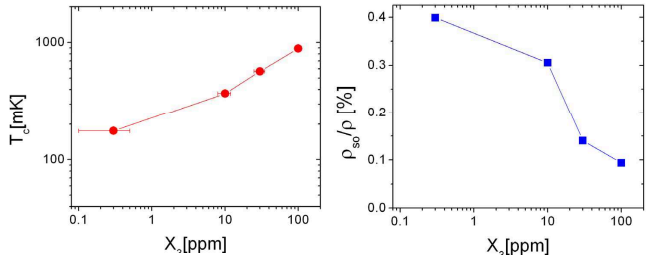}
\end{center}
\vskip -.7in \caption{Dependence of transition temperature and
superfluid stiffness on $^3$He concentration. Data of Chan et.\
al.\
\cite{Chan:05}, also quoted in \cite{Chan04}}.  \label{FIG:Fig3}
\vskip -.2in\end{figure}

Let us therefore address the effect of disorder on the supersolid
within the GL framework. In particular, because of its
experimental relevance, we are interested in the effect of $^3$He
impurities on the superfluid properties. One might expect that
disorder would suppress the supersolid state as it would interfere
with the phase ordering as it does in superconductors. However if
the supersolid arises due to the presence of vacancies, then the
presence of defects that {\em create} vacancies or that can
increase the local concentration of vacancies could lead to
enhanced superfluidity. Here, we offer some
speculations on the effects of a non-zero concentration of $^3$He
atoms on the transition temperature and superfluid stiffness.

1)  The presence of lighter $^3$He atom defects increases the zero
point fluctuations locally; this repels nearby host atoms. As we
have seen, a repulsion for $^4$He atoms is an attraction for
vacancies; this increases the local vacancy density as seen in
Fig.\ 2.
\begin{figure}[b]
\begin{center}
\vskip -.2in\includegraphics[width=8.2cm]{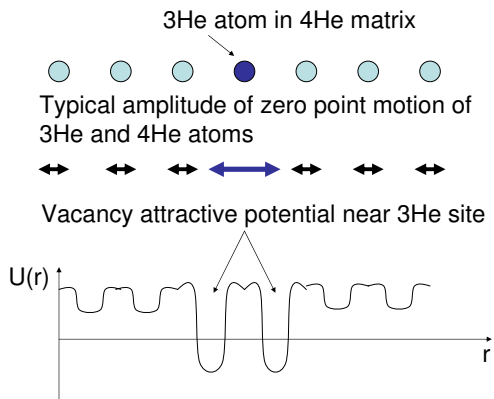}
\end{center}
\vskip -.4in
\caption{When included in the $^4$He lattice, $^3$He atoms  will
push the host atoms away. This redistribution is equivalent to an
attractive potential for vacancies, thus generating a larger
vacancy density near $^3$He.}
\label{FIG:Fig2}
\end{figure} This leads to an increase in $T_c$. To first
approximation, the effect would be linear in the $^3$He
concentration.  2) At the same time, there would be an effect on
the superfluid stiffness $\rho_s$. We suggest that
$\rho_s$ would decrease with increased $^3$He concentration
because $^3$He sites would produce attractive sites for the
superfluid condensate and localize it more. While $^3$He creates
puddles of enhanced supersolid condensate, the stiffness as a
reflection of the global phase rigidity would drop on average with
increased $^3$He doping. In other words,  to the extent that
substitutional impurities inhibit the exchanges of $^4$He atoms
that are required to enable vacancy hopping, we expect that any
foreign atom substitution will diminish the effective $t$, leading
to a reduction in superfluid stiffness. Again, the effect will be
linear in impurity concentration, see Fig(\ref{FIG:Fig3}).

So we propose that a $^3$He site is attractive for vacancies and
we model the effect of $^3$He substitution as a random attractive
potential for vacancy density with the potential  $u(\br)/\Omega =
+ \lambda\rho_3(\br)$,
where
$\lambda\, (>0)$ is the coupling strength of the effective $^3$He
--
$^4$He repulsion.  The corresponding contribution to the GL free
energy density  is
\beqa F_{3} =  - \lambda n_{3}|\psi_V(\br)|^2.
\eeqa Here, we have taken an annealed average,
$\langle\rho_{3}(\br)\rangle = n_3$ as the average density of
$^3$He atoms.

We see, from Eqs.\ (12-14), that the effect of $^3$He substitution
is to increase the transition temperature from $T_c^0$ to \beqa T_c
= (1+ {\lambda\over a}n_3)T_c^0, \;\;\; \delta T_c/T^0_c = {\lambda
\,\delta n_3\over a} \eeqa an effect that is linear in $^3$He
concentration $n_3$. One could estimate the parameters by examining
the experimental data. In fact the numbers from  Chan et al
\cite{Chan:05} show an enormous effect. From the data, Fig.\
\ref{FIG:Fig2},  we find that shift $\delta T_c = 150$ mK, or
$\delta T_c/T_c \approx 0.75$ for an increase in $^3$He
concentration $\delta n_3$ of about 10 ppm. Put differently, the as
defined critical temperature nearly doubled upon adding $0.001\%$
impurities. To date, no known  superfluid states
exhibit such a stong sensitivity to
impurities.  This is one of the many puzzles in this system
\cite{foot}. 

Now we turn to stiffness corrections. We argued above that the
stiffness term will  soften in the presence of $^3$He defects. The
effective hopping $t$ in the GL free energy density of Eq.\ (13)
is then $t = t_0(1 - g \, n_3/n_V)$ with
$g>0$. Since $\rho_s \propto t|\psi_V|^2$, we conclude that
 \beqa
  \rho_s = \rho^0_s(1- g\, n_3/n_V), \ \   \rho^0_s = 2( t_0
/a_0)\  |\psi_V|^2.
 \eeqa From the data, we estimate $\delta \rho_s/\rho_s^0 \approx
1/4$ for
$\delta n_3 \approx 10$ ppm. Assuming $n_V \approx 0.1\%$, we find
$g\approx 25$.

\section{Conclusion}

Although we motivated  our arguments with  a wave function
appropriate for considerations in the grand canonical ensemble, in
our discussion of the effect of $^3$He impurities on the superfluid
stiffness, we assumed implicitly that changes in the average
concentration of vacancies $n_V$ with $^3$He concentration are much
less important than the decrease in the kinetic energy. We have no
evidence for this but strictly speaking we cannot exclude the
possibility that adding $^3$He increases the equilibrium vacancy
density. If this is the case, we need to augment our arguments by
explicitly allowing $n_V$ be dependent on $n_3$. This would be an
extension of our analysis that could lead to modifications of our
results. Such an increase in $n_V$ would be a simple source of the
$T_c$ increase with addition of $^3$He. However, even the presence of patches
of higher $T_c$ could increase the bulk $T_c$ by proximity effects.
Similarly, the mass decoupling that is measured in the torsional
oscillator would have higher onset temperature due to patches of
higher $T_c$. We leave this issue until the experimental situation
is clearer.

We presented a GL approach to the
supersolid phase  that contains two order parameters, density
modulation
$n_B(\br)$ and superfluid
$\psi_V(\br)$.  Both the solid and superfluid fields are of course
made from the same $^4$He atoms and realize a dual quantum
mechanical behavior of these atoms. As others have argued, it
seems reasonable to us that lattice defects, for example
vacancies, are crucial for supersolid phenomena.

We find that disorder from  $^3$He atoms produces regions of
depleted $^4$He density, thus attracting vacancies. This
attractive potential due to disorder can be though of as creating
locally regions of higher $T_c$ in the GL functional for
$\psi_V$.  At the same time the granularity in $\psi_V$ produced
 by disorder leads to suppression of superfluid stiffness.
 Both of these effects are linear in concentration of $^3$He atoms.

This work was supported by DOE LDRD at Los Alamos and  by the Center
for Materials Theory at Rutgers  University. We thank M.W.H. Chan
for permission to use his figures and we acknowledge fruitful
discussions with  C. Batista, S. Trugman, P.W.\ Anderson, W.F.\
Brinkman, D.A.\ Huse,  R.\ Shankar, and A.E.\ Ruckenstein.


\begin{thebibliography}{99}
\bibitem{activ} Furthermore, we
believe that the activation energy for interstitials is larger
than that for vacancies and the bandwidth of the former is likely
to be smaller than that of the latter.
\bibitem{multi} We are aware that simple nearest neighbor hopping
may not be adequate to describe the multiparticle exchanges
responsible for vacancy motion.
\bibitem{KE} The kinetic energy operator in the spin
language is
$(-t/2)\sum [\sigma_i\cdot\sigma_j -\sigma_i^z \sigma_j^z].$
\bibitem{foot}According to Anderson, Brinkman and Huse
\cite{ABH}, the
$T_c$ determined in the experiments is actually not a true
transition temperature but some vortex dynamics temperature and
that the true $T_c$ cannot be determined directly from the
torsional oscillator experiments. If this objection is correct
then we cannot use existing data to estimate the $\lambda$
coefficient. Nevertheless the prediction about $T_c$ increase
would be correct.
\end{thebibliography}
\end{document}